\documentclass[aip,apl,reprint]{revtex4-2}

\usepackage{graphicx}
\usepackage{dcolumn}
\usepackage{bm}
\usepackage{amsmath}
\usepackage{multirow}
\usepackage{array}
\usepackage{comment}

\newcommand{\mum}{\mu\mathrm{m}}

\begin{document}

\title{A Phononic Crystal Waveguide Using Surface Waves Below the Sound Cone}

\author{Karanpreet Singh}
\author{Gabe Wilson}
\author{James A.H. Stotz}
\email{jstotz@queensu.ca}

\affiliation{Department of Physics, Engineering Physics \& Astronomy,\\ Queen's University
Kingston, ON  K7L 3N6, Canada.}

\date{\today}

\begin{abstract}

Surface acoustic waves are commonly used in a variety of radio-frequency electrical devices as a result of their operation at high frequencies and robust nature.  For devices based on Rayleigh-like plane waves, functionality is based on the fact that the Rayleigh wave mode is confined at the solid-air interface.  However, to create advanced functionality through the use of phononic crystal structures, standard cylindrical inclusions have been shown to couple Rayleigh modes to the shear horizontal bulk modes and provide a significant pathway to energy loss. We introduce alternative inclusion shapes with a reduced 2-fold symmetry that lowers the speed of the Rayleigh-like surface acoustic wave to below that of the shear horizontal mode.  With an eigenfrequency below the sound line, the new mode is confined to the surface with limited coupling and loss to the bulk.  Based on these inclusions, an acoustic waveguide design is proposed that demonstrates a strong confinement of wave energy both at the surface and within the waveguide.

\end{abstract}


\maketitle 


Surface acoustic waves (SAWs) are essential in diverse applications \cite{Delsing19}, from high-frequency filtering in telecommunications \cite{lam16} to precise sensors \cite{ding12, ding13, mujahid17, Mandal22} and advanced quantum computing \cite{Barnes00, bienfait19} functionalities. However, previous SAW waveguide designs using phononic crystals (PnCs) \cite{Laude15,Vasileiadis21} have primarily relied on line defects \cite{Khelif04, Benchabane15, Modica20, Laude21}, but can suffer from high propagation loss due to backscattering and leakage into the surrounding PnC, which hampers their efficiency and restricts miniaturization or formation of complex acoustic circuits. Although topological materials have been explored to address these issues \cite{yan-feng21, yan-feng22, friend2023}, practical integration challenges and fabrication constraints persist.

An alternative to creating a line-defect waveguide in a PnC on a bulk substrate is to alter the surface properties to create a slow channel through which the wave can propagate.  Previous work includes the use of pillars\cite{AlLethawe16} or inclusions\cite{Muzar23} to modify the surface velocity, and the surrounding, unmodified surface naturally acts as the cladding and eliminates the need for separate, extended cladding structures.  Challenges do persist in these structures as care must be taken to not have modes that exist above the sound line\cite{Benchabane06}, where coupling to radiative bulk modes leaks energy away from the surface and into the bulk.

Focusing on waveguides using inclusions to create a PnC structure, dispersion calculations for a square lattice of shallow inclusions in GaAs showed that the velocity of a Rayleigh-like SAW mode is smaller than the pure Rayleigh mode of a bare surface\cite{Muzar19}.  A PnC waveguide was demonstrated\cite{Muzar23} that capitalized on the reduced SAW velocity, but it was evident that inclusions coupled the Rayleigh mode to the previously orthogonal shear horizontal (SH) bulk mode.  For such inclusions on the surface, the sound line threshold effectively becomes the shear horizontal mode.  This is supported by dispersion calculations of shallow inclusions\cite{Muzar19}, which show a dramatic reduction in loss to the bulk when the eigenfrequencies of the Rayleigh-like mode of a PnC fell below that of the SH mode.  In that work, this occurred for modes near the boundary of the Brillouin zone, which are less practical as the group velocity tends to zero. Therefore, reducing the eigenfrequencies of Rayleigh-like SAW modes (and, equivalently, the phase velocity of those modes) to below that of the SH mode brings those SAW modes below the sound line, eliminates radiation pathways of acoustic energy to the bulk, and maintains the surface confinement of the SAWs in the microstructured surface.


\begin{figure}[t]
    \centering
    \includegraphics[width=\linewidth]{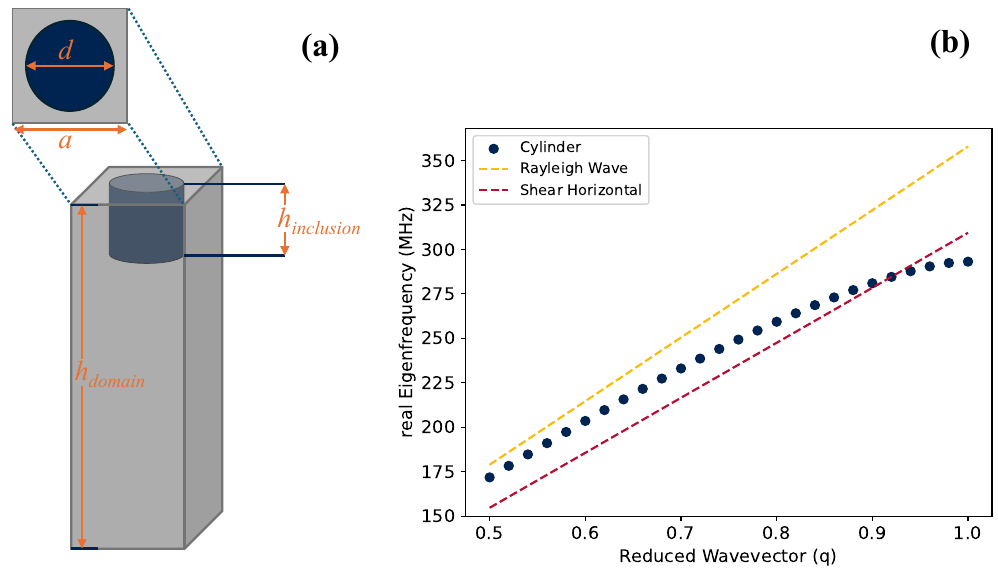}
    \caption{(a) Visualization of the single unit cell domain. For this study $a = ~4~\mum$, $h_{inclusion} = 3~\mum$ and $h_{domain} = 50~\mum$ while $d$ varies between shapes (b) Dispersion relation comparing the eigenfrequency of the Rayleigh mode in an infinite crystal of cylindrical inclusions to that of the Rayleigh mode on a bare surface and bulk shear horizontal waves.}
    \label{fig:cylinder_vs_bare_Band_Diagram}
\end{figure}


In this Letter, previous PnC designs are modified by exploring alternative inclusion geometries to overcome the limitations imposed by the SH mode. The shape of inclusions on PnCs is important\cite{Paul24}. In particular, the current work demonstrates that by changing the inclusion design from simple cylinders to elongated shapes, the SAW eigenfrequency is sufficiently reduced to less than that of the SH mode thus achieving improved waveguide confinement in both the lateral and vertical directions. Both a rectangular prism and elliptical cylinder inclusions are introduced to contrast with the conventional circular cylinder inclusion.  While both new shapes improve upon traditional cylindrical shapes, it is the elliptic cylinder that lowers the eigenfrequency of the its modes the most, resulting in a waveguide that possesses the superior lateral confinement to previous work while improving the surface confinement figure of merit by ten orders of magnitude. 


The phononic crystal waveguides were designed and simulated in COMSOL Multiphysics along with the Structural Mechanics module, which uses finite element methods for eigenfrequency analysis.  Results provided eigenfrequencies and corresponding eigenmodes of the various inclusion and domain geometries under investigation.  The lattice parameter ($a$) of all PnCs was set to $4~\mum$, and the depth of the inclusions (holes) was fixed at $3~\mum$. The stiffness constant and density for GaAs were taken from Tanaka and Tamura \cite{Tanaka98} and the COMSOL Material Library, respectively, which are presented in Table \ref{tab:material_constants}. The wave propagation is along the $\Gamma$-$X$ direction which is oriented along the crystallographic [110]-direction (or equivalent) of GaAs, which is the the piezoelectric direction of a GaAs (001)-surface where SAWs may be excited.  To accommodate SAWs traveling along the [110]-direction, the materials parameters were rotated using a bond transformation matrix approach \cite{NorouzianTurner}. 

To characterize the effects of elongated inclusions, two types of analysis are provided that, by extension, require two computational domains:

\begin{enumerate}

    \item \textbf{Dispersion Relation:} The dispersion relation of various inclusions shapes (cylindrical, rectangular prism, and elliptic cylinder) are modeled within a single unit cell domain with dimensions $4~\mum \times 4~\mum \times 50~\mum$, as shown in Fig.~\ref{fig:cylinder_vs_bare_Band_Diagram}. This geometry models an infinite PnC and is used to identify geometries and Rayleigh-like modes that have eigenfrequencies below the bulk SH mode.
    
    \item \textbf{Waveguide Design:} Using a suitable inclusion shape and acoustic wavevector, four inclusions are arranged side-by-side in a larger computational domain to form a waveguide $16~\mum$ wide. The entire waveguide structure was simulated on computational domain of $4~\mum \times 100~\mum \times 50~\mum$, and the properties of the Rayleigh-like eigenmode are evaluated.
    
\end{enumerate}


\begin{table}[t]
    \centering
    \caption{\label{tab:material_constants} Material Constants of GaAs}
    \renewcommand{\arraystretch}{1.2} 
    \begin{ruledtabular}
    \begin{tabular}{>{\centering\arraybackslash}m{0.55\linewidth} >{\centering\arraybackslash}m{0.35\linewidth}}
        Material Constant & Value \\
        \hline
        Stiffness Constant, $c_{11}$ & 1.19$\times 10^{11}$ Pa \\
        Stiffness Constant, $c_{12}$ & 5.38$\times 10^{10}$ Pa \\
        Stiffness Constant, $c_{44}$ & 5.94$\times 10^{10}$ Pa \\
        Density, $\rho$ & 5307~kg/m$^3$ \\
    \end{tabular}
    \end{ruledtabular}
\end{table}


Appropriate boundary conditions were chosen to accurately represent the physical system. The top surface was modeled as a free boundary in both computational domains, allowing SAWs to propagate freely. For the single unit cell domain used for dispersion relation calculations, opposite vertical sides were paired using Bloch-Floquet periodic boundary conditions to simulate an infinite periodic array. In the waveguiding domain, Continuity periodic boundary conditions were employed on the side walls perpendicular to wave propagation.  This helps to ensure displacement and stress continuity as edge effects are minimal. The boundary conditions remained Bloch-Floquet in the direction of wave propagation.  In both domain geometries, a Low-reflecting boundary was applied to the bottom surface to minimize reflections and simulate a semi-infinite substrate.\cite{Muzar23} Further, a mesh sensitivity analysis was performed to balance accuracy and computational efficiency, resulting in the selection of a maximum mesh size of $a/4$ ($1~\mum$) and a minimum mesh size of $a/40$ ($0.1~\mum$), which is comparable to previous work.\cite{Muzar23}

To focus the study to Rayleigh-like SAW eigenmodes, three key parameters were used. The first is the squared polarization ratio ($r$), which is defined as the ratio of the displacement components in the $xz$-plane (sagittal plane) to the total displacement, and correlates to the polarization of a Rayleigh wave.  The ratio is defined as 

\begin{equation}
    r = \frac{\int(u_x^*u_x + u_z^*u_z)dV}{\int(u_x^*u_x + u_y^*u_y + u_z^*u_z)dV},
    \label{first_equation}
\end{equation}

\noindent and is equal to 1 for an plane Rayleigh wave.  The second parameter used is the strain energy ratio.  This is calculated by dividing the integrated elastic strain energy in the upper half of the domain for a particular eigenmode by the elastic strain energy in the entire domain.  The strain energy ratio is a useful measure of identifying whether a mode has most of its energy at or near the surface as would a SAW.  By considering eigenmodes where the value of both the squared polarization ratio and the strain energy ratio are greater than 0.6, SAWs can be readily identified in the PnC structure that are Rayleigh-like.  For the purposes of this discussion, we will only be considering the lowest frequency band in the dispersion relation as the focus is to identify modes with frequencies below the bulk SH mode. Finally, we also introduce the parameter referred to as the logarithmic reciprocal attenuation (or, simply, the reciprocal attenuation), which will serve as our figure of merit in the discussion below.  It is defined as $-\log_{10}\left(\frac{\text{Im}(\omega)}{\text{Re}(\omega)}\right)$ and corroborates well with the strain energy ratio, quantifying precisely the attenuation of the eigenmodes as energy is lost to the bulk, as modeled by the absorbing boundary on the bottom of the simulation domain.



\begin{figure}[t]
\centering
\includegraphics[width=\linewidth]{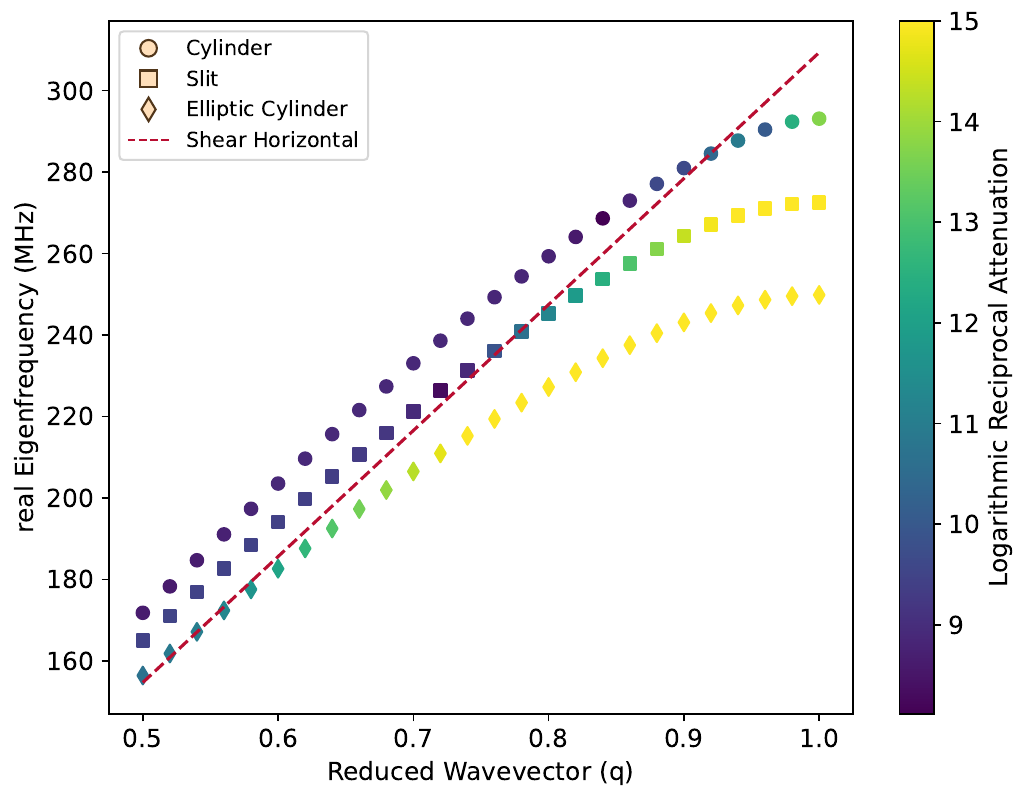}
\caption{Dispersion relation comparing the lowest order SAW modes of cylindrical, rectangular prism, and elliptical cylinder inclusions to the shear horizontal threshold.}
\label{fig:multiple_shapes_Band_Diagram}
\end{figure}


To investigate the impact of inclusion shape, the dispersion relations of the lowest Rayleigh-like mode are compared for three inclusion shapes:  cylindrical, rectangular prism, and elliptical cylinder.  Each inclusion is centered on the surface of a single unit cell of the domain and is designed to maintain a consistent dimension of 3.4 microns along the major axis. Specifically, the cylindrical inclusion has a diameter of $3.4~\mum$, and both the rectangular prism and elliptical cylinder maintain an aspect ratio of 3:1. For the elliptical cylinder, this yields a semi-minor axis of $0.566~\mum$ and a semi-major axis of $1.7~\mum$.  Similarly, the slit has dimensions of $3.4~\mum \times 1.133~\mum$. In both cases, the major axes are perpendicular to the direction of wave propagation. As mentioned above, the depth of all inclusions is maintained at $3~\mum$ across all shapes.

Figure~\ref{fig:multiple_shapes_Band_Diagram} presents the dispersion relation for these three inclusion geometries, plotting eigenfrequency against the reduced wavevector $q = k/k_x$, where $k$ is the wavevector of the SAW and $k_x = \frac{\pi}{a}$ is the boundary of the Brillouin zone for the square lattice of inclusions. The dashed red line represents the SH wave eigenfrequencies of bulk GaAs, serving as the critical threshold for decoupling SAWs from bulk modes. As evident from Fig.~\ref{fig:multiple_shapes_Band_Diagram}, elliptical cylinder inclusions achieve eigenfrequencies consistently below the SH threshold for wavevectors well displaced from the boundary of the Brillouin zone.  As this mode diverges from the SH line, the reciprocal attenuation noticeably increases indicating that coupling to the bulk modes is dramatically diminished, which further implies a robust confinement of the SAW to the surface even in the presence of inclusions.  The strain energy ratio and squared polarization ratio parameters, are consistently high across all $q$ values.  The strain energy ratio $> 0.9998$ and squared polarization ratio ranging from 0.9810 to 0.9845) further indicating that these are surface bound Rayleigh-like waves.

The elliptical cylinder provides two important aspects that slow the phase velocity of the SAW.  Like a typical cylindrical inclusion, the elliptical cylinder has local resonant modes that are excited as the wave propagates on the surface.  Moreover, as the cylindrical inclusion transforms into the 2-fold symmetric ellipse shape, the volume of the inclusion reduces.  This effectively increases the mass at the surface, and this mass-loading decreases the eigenfrequencies of the SAW mode.  For comparison, the rectangular prism inclusion is also included in Fig.~\ref{fig:multiple_shapes_Band_Diagram}, which has the same maximal dimensions as the elliptical cylinder.  This 2-fold symmetric inclusion shape also reduces its eigenmode frequencies below the SH mode before the boundary of the Brillouin zone, but not to the same extent as the elliptical cylinder.  This again supports the concept of mass-loading effects, as the volume of rectangular prism inclusion would be larger than that of the elliptical inclusion.  

\begin{figure}[t]
    \centering
    \includegraphics[width=\linewidth]{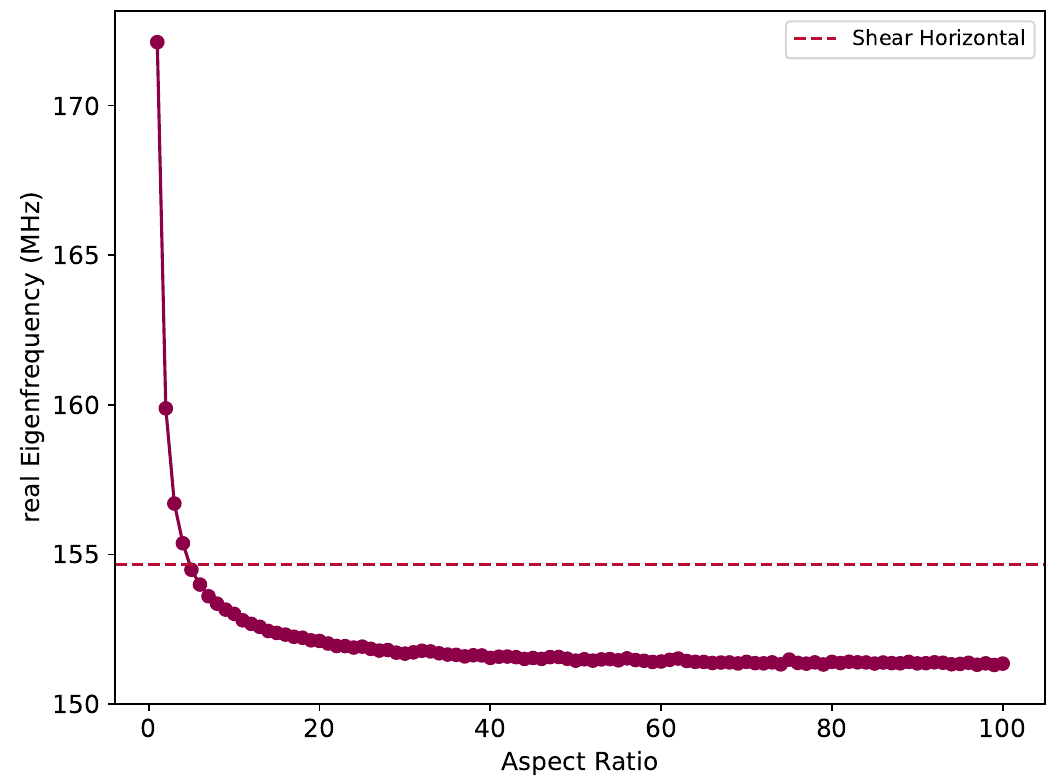}
    \caption{Rayleigh mode eigenfrequency at $q=0.5$ as a function of the aspect ratio (major to minor axis) of the elliptical cylinder inclusion.  The first point presents an aspect ratio of 1, which is the cylindrical inclusion. The frequency of the SH mode is represented by the dotted line.}
    \label{fig:aspect_ratio}
\end{figure}

To further explore the effect of mass loading, Fig.~\ref{fig:aspect_ratio} plots the eigenfrequency for modes at $q=0.5$ for the surface profile of the inclusions transitioning from a circle to a high aspect ratio ellipse.  As the aspect ratio increases, there is a monotonic decrease in the eigenfrequency, and as the shape adopts its 2-fold degeneracy, there is a rapid change from lower aspect ratios.  This is the direct result of a significant addition of mass added around the inclusion as the aspect ratio transitions from 1:1 to 4:1. At higher aspect ratios, the eigenfrequency drops below the SH mode, and the changes eventually become muted as the incremental mass loading is significantly reduced.  It is interesting that while the minor axis of the highest ratios is incredibly small (on the order of 30~nm), the presence of the discontinuity in the surface continues to govern the physics of the PnC as a local resonator.   


The above clearly underscores the elliptical cylinder’s effectiveness for reducing the eigenfrequencies of the SAW and decoupling the surface mode from bulk radiation pathways, and the inclusion is suitable to be used as a basis for phononic crystal-based waveguides.  To compare with previous work\cite{Muzar23}, a simulation domain is constructed employing a waveguide consisting of four elliptical cylinders side by side, forming a $16~\mum$ wide waveguide structure with the unmodified GaAs surface as the cladding.  The elliptical cylinders were chosen to have the 3:1 aspect ratio to align with readily available optical lithography techniques. To quantify energy confinement within the waveguide core, we introduced a new parameter, the core strain energy, which represents the fraction of strain energy confined in the unit cells below the waveguide inclusion and within the upper $25~\mum$ of the computational domain. The  core strain energy ratio is then calculated by dividing this core strain energy by the total strain energy in the computational domain.


\begin{figure}[t]
    \centering
    \includegraphics[width=\linewidth]{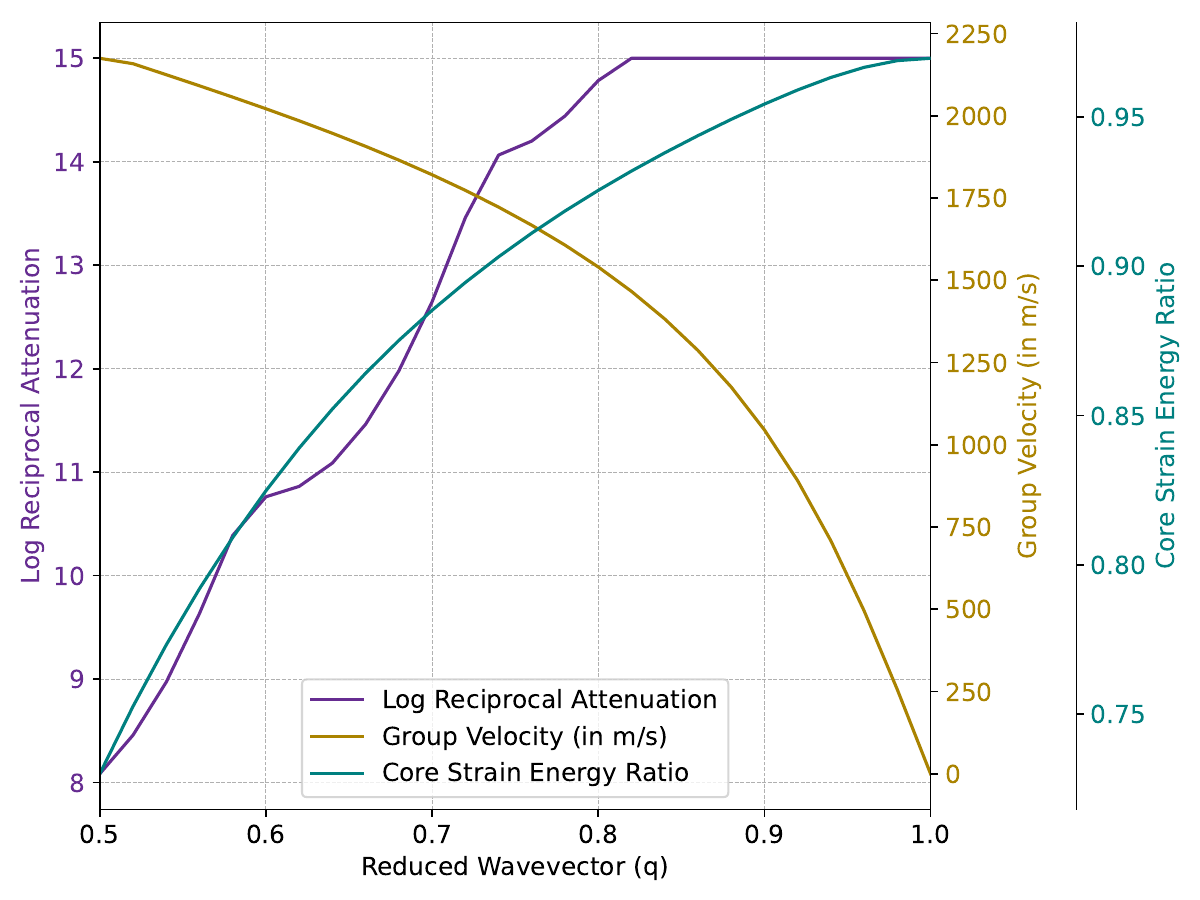}
    \caption{Waveguide performance diagram showing key metrics as functions of the reduced wavevector $q$ for the four-inclusion waveguide with elliptical cylinders.}
    \label{fig:waveguide_performance}
\end{figure}


Figure~\ref{fig:waveguide_performance} presents the waveguide performance diagram, plotting key parameters —- reciprocal attenuation, group velocity, and core strain energy ratio —- against the reduced wavevector $q$ for the Rayleigh-like eigenmodes propagating through the waveguide.  As observed for the infinite crystal in Fig.~\ref{fig:multiple_shapes_Band_Diagram}, the reciprocal attenuation is monotonically increasing as $q$ increases, or similarly, as the acoustic wavelength decreases.  With a decrease in wavelength, the SAW will naturally be more situated near the surface, which enhances the effect of the PnC on wave motion.  Interestingly, near $q=0.6$, the slope of the increase changes, which is about where the PnC eigenmode crosses the SH mode in the dispersion relation.  An additional change in slope is observed near $q=0.74$, which could indicate the presence of an additional coupling mechanism as the eigenfrequencies drops further below the SH sound line.  As $q$ approaches 1, the reciprocal attenuation saturates, which suggests that a limit has been reached in COMSOL when the imaginary frequency is very small.  Also indicated in Fig.~\ref{fig:multiple_shapes_Band_Diagram}, the group velocity is naturally decreasing as $q$ approaches the boundary of the Brillouin zone.  As a result of this and the increasing reciprocal attenuation, the core strain energy ratio increases for increasing $q$ as more acoustic energy is confined both within the waveguide and near the surface.

A waveguide operated at $q = 0.74$ provides a balanced approach from the performance metrics: the group velocity is $1722 m/s$, such that the wave is adequately slowed for effective waveguiding while still enabling acoustic transport; the core strain energy ratio is large ($0.903$) indicating that over $90\%$ of the total energy is stored within the waveguide core; and the reciprocal attenuation is $10^{-14}$, demonstrating exceptional SAW confinement to the surface. Notably, this reciprocal attenuation value is almost ten orders of magnitude higher than that achieved with cylindrical inclusions in previous designs\cite{Muzar23}.


\begin{figure}[t] 
    \centering
    \includegraphics[width=0.5\textwidth]{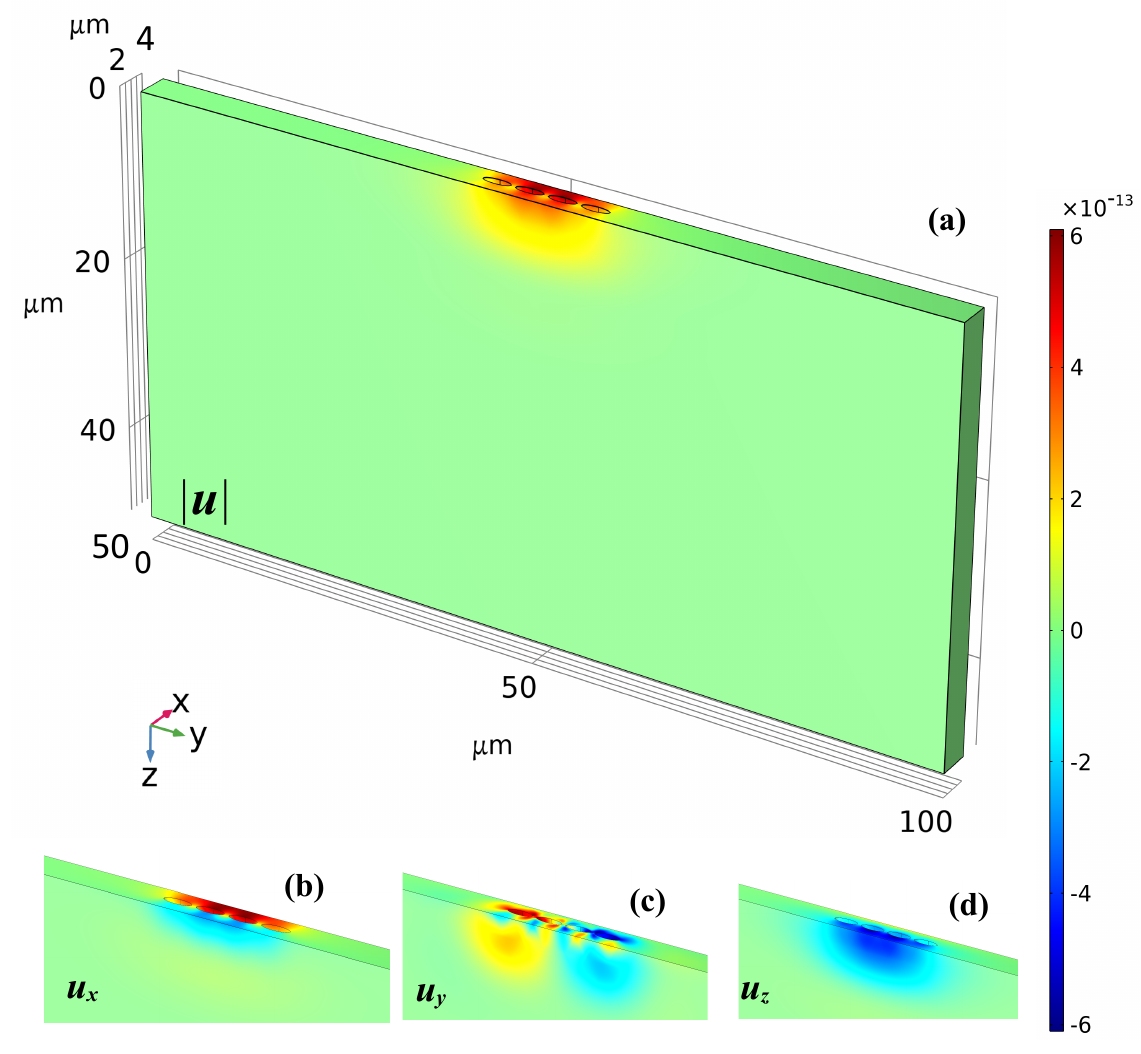}
    \caption{Magnitude of Displacement Field across a 4-inclusion waveguide using elliptical inclusions for $q=0.74$ and: (a) total displacement; (b) \textit{x-}component; (c) \textit{y-}component; (d) \textit{z-}component.  The displacement magnitudes for the \textit{x-} and \textit{z-}directions are comparable to the total displacement, but displacement in the \textit{y-}direction is an order of magnitude less.}
    \label{fig:waveguide_displacement_field}
\end{figure}


Figure~\ref{fig:waveguide_displacement_field} visualizes the displacement field within the four-hole waveguide of elliptical cylinder inclusions at $q = 0.74$. The individual components $u_x$, $u_y$, $u_z$ aid in visualizing the field distribution, while the norm of the displacement vector, $|u|$, serves as a quantitative measure of overall SAW confinement. Panel (a) shows the displacement magnitude norm ($|u|$), which is concentrated near the surface within the waveguide core, indicating effective lateral and vertical confinement of the surface acoustic waves. The high-intensity area localized within the waveguide core demonstrates strong confinement, preventing energy leakage into the bulk substrate, and predicts efficient acoustic waveguiding. An animation of the wave exhibiting all phases is included in the supplemental material.

Panels (b), (c), and (d) display the displacement components $u_x$, $u_y$, and $u_z$, respectively. The $u_x$ and $u_z$ components, corresponding to motion in the sagittal plane as desired for Rayleigh-like modes, and they exhibit symmetric profiles about the waveguide core.  In contrast, the $u_y$ component is an order of magnitude weaker, and is coupled to the sagittal motion via the elliptical inclusions.  This lateral motion shows an interesting anti-symmetric profile, but remains confined to the waveguide core in contrast to cylindrical inclusions\cite{Muzar23}, which extended well beyond the waveguide.

For every measure, the elliptical cylinder geometry excels in achieving significant confinement of acoustic energy to the surface and lateral confinement within a simple waveguide.  This is made possible because of the targeted lowering of the SAW eigenfrequency below the shear horizontal threshold using the 2-fold symmetric elliptical cylinder.  As a result, the desired surface modes are below the sound line, which now includes the shear horizontal mode, and efficient phononic crystal waveguides can be made on the surface of a substrate.  In particular, the design enhanced the surface confinement of a waveguide by almost 10 orders of magnitude using the reciprocal attenuation as a figure of merit.  Such a system can also form a new platform on which PnC structures and waveguides are applied.  For instance, this configuration is highly suited for applications in quantum information processing and spintronics \cite{Helgers22,Stotz05,Stotz08}, where laterally confined SAWs are critical for the coherent transport of spin information.

\ 

The authors would also like to gratefully recognize CMC Microsystems for the provision of products and services that facilitated this research, which includes the use of the COMSOL Design Tool.



%

\end{document}